\pdfoutput=1

\documentclass[twocolumn,
amsmath,amssymb,
aps,prl,reprint, nofootinbib, nobibnotes,
superscriptaddress
]{revtex4-1}

\usepackage{subfigure}
\usepackage{graphicx}
\usepackage{dcolumn}
\usepackage{bm}
\usepackage{graphicx}
\usepackage[toc,page]{appendix}
\usepackage{blindtext}
\usepackage{physics}

\usepackage{pgfplots}
\pgfplotsset{compat=1.13}

\usepackage[version=4]{mhchem}

\usepackage{hyperref}

\usepackage[overlay,absolute]{textpos}
\newcommand\PlaceText[3]{
	\begin{textblock*}{10in}(#1,#2)  
		#3
	\end{textblock*}
}

\begin{document}
	
\title{Robust two-qubit gates in a linear ion crystal using a frequency-modulated driving force}

\author{Pak Hong Leung}
\email{pleung6@gatech.edu}
\affiliation{School of Physics, Georgia Institute of Technology, Atlanta, Georgia 30332, USA}
\author{Kevin A. Landsman}
\author{Caroline Figgatt}
\author{Norbert M. Linke}
\author{Christopher Monroe}
\affiliation{Joint Quantum Institute and Joint Center for Quantum Information and Computer Science, University of Maryland, College Park, MD  20742,  USA}
\author{Kenneth R. Brown}
\affiliation{School of Physics, Georgia Institute of Technology, Atlanta, Georgia 30332, USA}
\affiliation{Schools of Chemistry and Biochemistry and Computational Science and Engineering, Georgia Institute of Technology, Atlanta, Georgia 30332, USA}

\date{\today}

\begin{abstract}
	In an ion trap quantum computer, collective motional modes are used to entangle two or more qubits in order to execute multi-qubit logical gates. Any residual entanglement between the internal and motional states of the ions results in loss of fidelity, especially when there are many spectator ions in the crystal. We propose using a frequency-modulated (FM) driving force to minimize such errors. In simulation, we obtained an optimized FM two-qubit gate that can suppress errors to less than 0.01\% and is robust against frequency drifts over $\pm$1 kHz. Experimentally, we have obtained a two-qubit gate fidelity of $98.3(4)\%$, a state-of-the-art result for two-qubit gates with 5 ions.
\end{abstract}

\maketitle

Ion traps are a leading candidate for the realization of a quantum computer. Magnetically insensitive qubit energy splittings, long coherence times, and high-fidelity state initialization and detection \cite{yb1, yb2} prove to be significant advantages for trapped ion qubits. Individual qubit addressing and single-qubit gates with error rates on the order of $10^{-5}$ per gate have been achieved \cite{yb1, single1, single2, microwave7}. Multiple qubits can be entangled through state-dependent forces driven by external fields \cite{Molmer, Sorensen, Milburn, Solano}, and for exactly two ions, entangling gate fidelities routinely exceed 99\% and in some cases 99.9\%. \cite{Lucas1, Lucas2, HF_Wineland, microwave1, microwave6, 14-qubit}. 

With increasing ion number, however, the motional modes bunch in frequency, which means exciting only a single motional mode becomes prohibitively slow. Alternatively, the state-dependent driving forces can couple to all modes of motion. A number of schemes have been proposed for disentangling the internal qubit states from the motional states of all modes by introducing variations to the driving force during the gate. One way to achieve this goal is amplitude modulation (AM) of the driving field \cite{AM1, Roos}. Several experiments have adopted this method and have achieved a 2 to 5\% error \cite{small_computer, spin-spin_coupling, optimal_control}. Discrete phase modulation (PM) has also been proposed for the same purpose, but the number of pulses in the sequence increases exponentially with the number of ions \cite{Phase_decoupling}. Moreover, discrete changes in laser amplitude and phase are hard to implement physically, especially when we perform fast gates.

We propose a novel decoupling method through continuous frequency modulation (FM), theoretically equivalent to continuous PM, which involves only small and smooth oscillations of the detuning of the applied field. First, we explain the coherent displacement of the ion chain's motional modes during the M\o lmer-S\o rensen (MS) gate. Then, we describe how the residual displacement of the ions can be minimized in a way which is robust to small changes in trap frequency. Next, we experimentally demonstrate this gate in a chain of 5 $\ce{^{171}Yb+}$ ions. Finally, we discuss extensions of the method to larger ion chains, with 17 ions as an example.

To entangle two qubits with the MS gate, we apply a state-dependent driving force near the sideband frequencies. As a result, each motional mode experiences a coherent displacement characterized by the operator\cite{AM1, Roos}: 

\begin{equation}
\begin{aligned}
\hat{D}(\hat{\alpha_k}) &= \exp(\hat{\alpha}_k a_k^\dag-\hat{\alpha}_k^\dag a_k),\\
\hat{\alpha}_k (t) &= \frac{\Omega}{2}(\eta_{i,k}\sigma_\phi^i+\eta_{j,k}\sigma_\phi^j)\int^t_0 e^{i\theta_k(t')} dt'
\end{aligned}
\end{equation}
\\
where $\Omega$ is the carrier coupling strength, $\eta_{i,k}$ and $\eta_{j,k}$ are the Lamb-Dicke parameters of ions i and j with respect to mode k, $\sigma_\phi^i$ and $\sigma_\phi^j$ are bit-flip Pauli operators for the addressed ions, and $\theta_k(t) = \int^t_0 \delta_k (t') dt'$ and $\delta_k (t)$ are the phase and detuning of the driving force relative to mode k. If the qubits are at the +1 eigenstate of both $\sigma_\phi^i$ and $\sigma_\phi^j$, the displacement is:

\begin{equation}
\alpha_k (t) = \frac{\Omega}{2}(\eta_{i,k}+\eta_{j,k})\int^t_0 e^{i\theta_k(t')} dt'
\end{equation}

We may visualize the trajectory of $\alpha_k(t)$ over time by plotting it in the complex plane. This is the phase space trajectory (PST) of the motional mode k. For a total gate time $\tau$, $\alpha_k(0) = 0$ and $\alpha_k(\tau)$ are the beginning and end points of the PST.

\begin{figure} [h]
	\centering
	\scalebox{.85}{\includegraphics{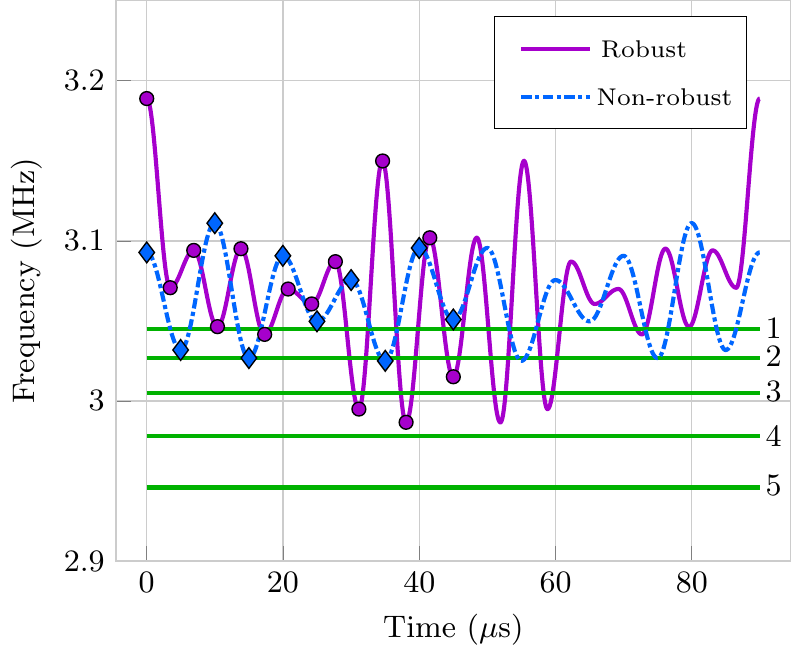}}
	\caption{Robust (violet, solid) and non-robust (blue, dash-dotted) FM pulses for 2-qubit gate optimized for 5 ions, both with a gate time of 90 \textrm{$\mu$}s. Green lines are experimental sideband frequencies, labeled 1 to 5, the first one being the common mode frequency. The pulses are designed to be symmetric in time. The dots and diamonds are the vertices of the frequency and represent the control parameters allowed to vary in our optimization algorithm.}
\end{figure}

Due to the state-dependent nature of $\hat{\alpha}_k (t)$, different eigenstates of $\sigma_\phi^i$ and $\sigma_\phi^j$ follow different PSTs. If any of the $\alpha_k (\tau)$ is non-zero, there is residual entanglement between the internal and motional state spaces, which leads to a mixed internal state. This lowers the overall gate fidelity ($F = |\langle\psi_{final}|\psi_{ideal}\rangle|^2$). Given that $|\alpha_k| \ll 1$, we find that the consequent gate error may be estimated as:

\begin{equation}
\varepsilon \equiv 1 - F \approx \sum\limits_{k=1}^{N}|\alpha_k(\tau)|^2
\end{equation}

Minimizing $|\alpha_k|$ is therefore the most straightforward criterion for an optimized gate. However, the gate is sensitive to small drifts in sideband frequencies ($\delta_k \rightarrow \delta_k + \delta_1$ and $\delta_1 \ll 1/\tau$), an imperfection which we often observe in experiments. The frequency dependence of $\alpha_{k}(\tau)$ can be canceled to the first order by setting the time-averaged position of $\alpha_{k}(t)$  to zero.

\begin{equation}
\alpha_{k,avg} \propto \int^{\tau}_{0} \int^{t}_{0}e^{i\theta_k (t')}dt'dt = 0
\end{equation}

It turns out that if we only consider symmetric pulses ($\delta_k(\tau - t) = \delta_k(t)$), minimizing $\alpha_{k,avg}$ also minimizes $\alpha_k(\tau)$. \smallskip

\begin{figure}
	\scalebox{0.3}{\includegraphics{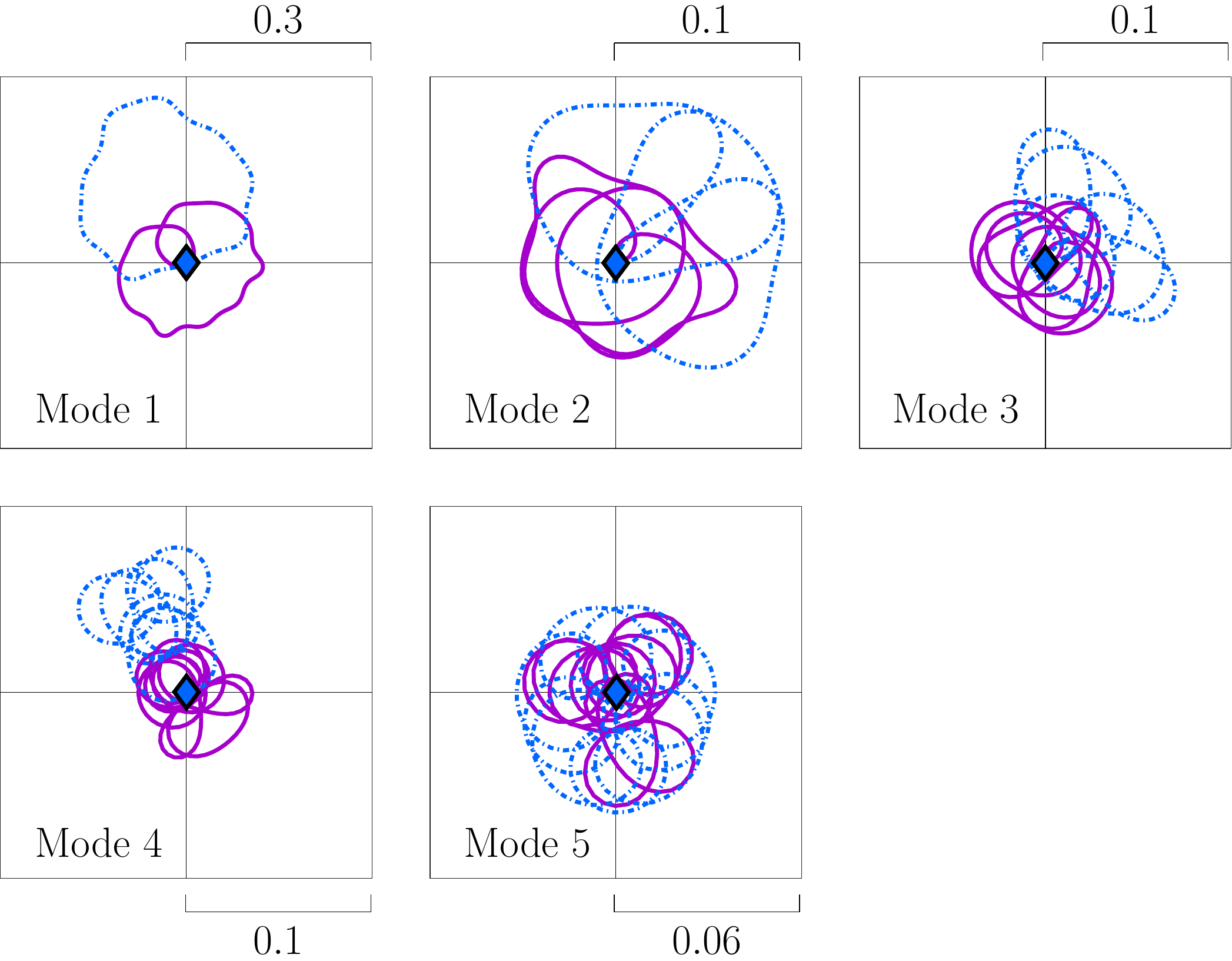}}
\end{figure}
\begin{figure}
	\scalebox{0.3}{\includegraphics{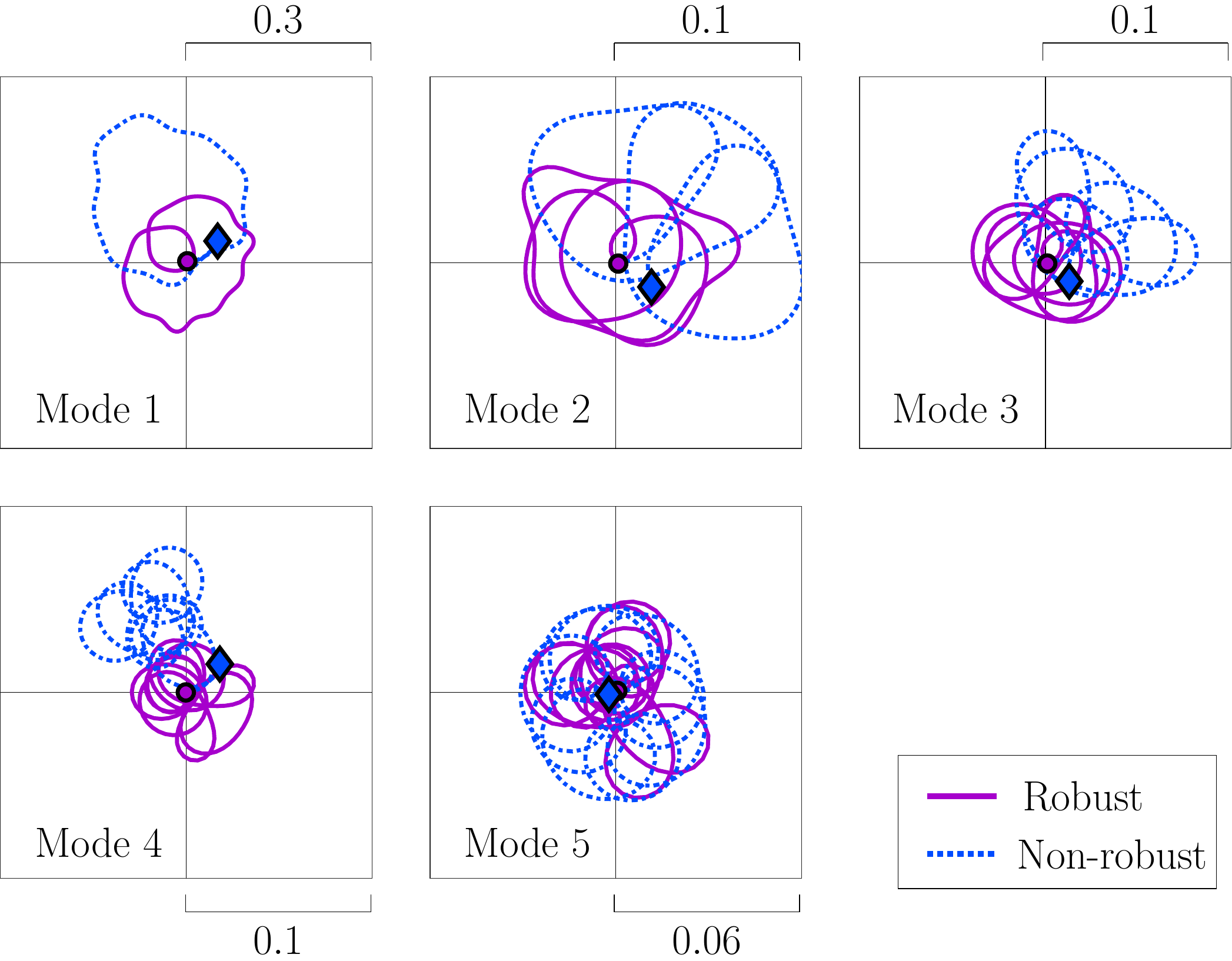}}
	\caption{\small Simulated PSTs with: (a) no frequency error and (b) -1 kHz sideband drift, using the FM pulses shown in Fig. 1. The end points for the robust pulse (circles) return to the starting point with the drift, whereas those for non-robust (diamonds) fail to do so. The horizontal and vertical axes represent the quadratures $x_k\sim a_k^\dag+a_k$ and $p_k\sim i(a_k^\dag-a_k)$ respectively.}
\end{figure}

\PlaceText{113mm}{20mm}{\small (a)}
\PlaceText{113mm}{75mm}{\small (b)}

In our scheme, we modulate the driving frequency during the gate to minimize the gate error. The trajectory $\alpha_k(t)$ moves with constant speed but varying angular rate $\delta_k(t)$. Therefore, FM allows us to control the curvature and thus the shapes and end points of the PSTs. We let the frequency assume a symmetric, oscillatory pattern (see example in Fig. 1). The vertices (local maxima and minima) of the oscillations are set to be evenly spaced in time and are the only variable control parameters in our optimization. The vertices are connected with sinusoidal functions, which leads to a smooth and continuous frequency profile. The function to be minimized is $|\alpha_{k,avg}|^2$ for robust pulses and $|\alpha_{k}|^2$ for non-robust. The number of vertices used is increased until we successfully converge to a solution with errors much lower than 0.01\%. Detailed derivations for equations (3) and (4) as well as the optimization process are provided in the Supplemental Material. 

Both robust and non-robust versions of the gate are tested on our 5-ion quantum computer. In our setup, 5 $\ce{^{171}Yb+}$ ions are held in an rf Paul trap with a radial trap frequency of $3.045$ MHz and an average ion separation of about 5 \textrm{$\mu$}m. Our qubit is defined by the ground hyperfine states $\ce{^{2}S_{1/2}},\ket{F=0}$ and $\ce{^{2}S_{1/2}},\ket{F=1}$ with an energy splitting of $2\pi\times12.642821$ GHz \cite{yb1}. Initially, all ions are cooled to close to the motional ground state ($\approx$ 0.1 phonons) and then optically pumped to the $\ket{0}$ state. Quantum gates are implemented using a beatnote generated by counter-propagating Raman laser beams that are capable of addressing any individual qubit \cite{small_computer}.

The 5 transverse motional sidebands are experimentally determined and used to find the optimal FM pulses for the 2-qubit gate. We increase the number of oscillations (degrees of freedom) for optimization until we find a pulse with low errors. With a fixed gate time of 90 \textrm{$\mu$}s, the optimized robust pulse consists of 13 oscillations, whereas the non-robust version has only 9 (Fig. 1). The driving frequency crosses the sidebands multiple times, which contrasts with other implementations of MS gates that avoid sideband resonance.

PSTs are plotted for no frequency error and for a 1 kHz frequency drift for both robust and non-robust pulses in Fig. 2. With the drift, the end points of the robust trajectory (circles) stick to the origin, whereas those of the non-robust (diamonds) deviate from the starting point, causing an estimated error of about 0.5\%. This proves the importance of the robustness criterion.
\begin{figure}
	\scalebox{0.28}{\includegraphics{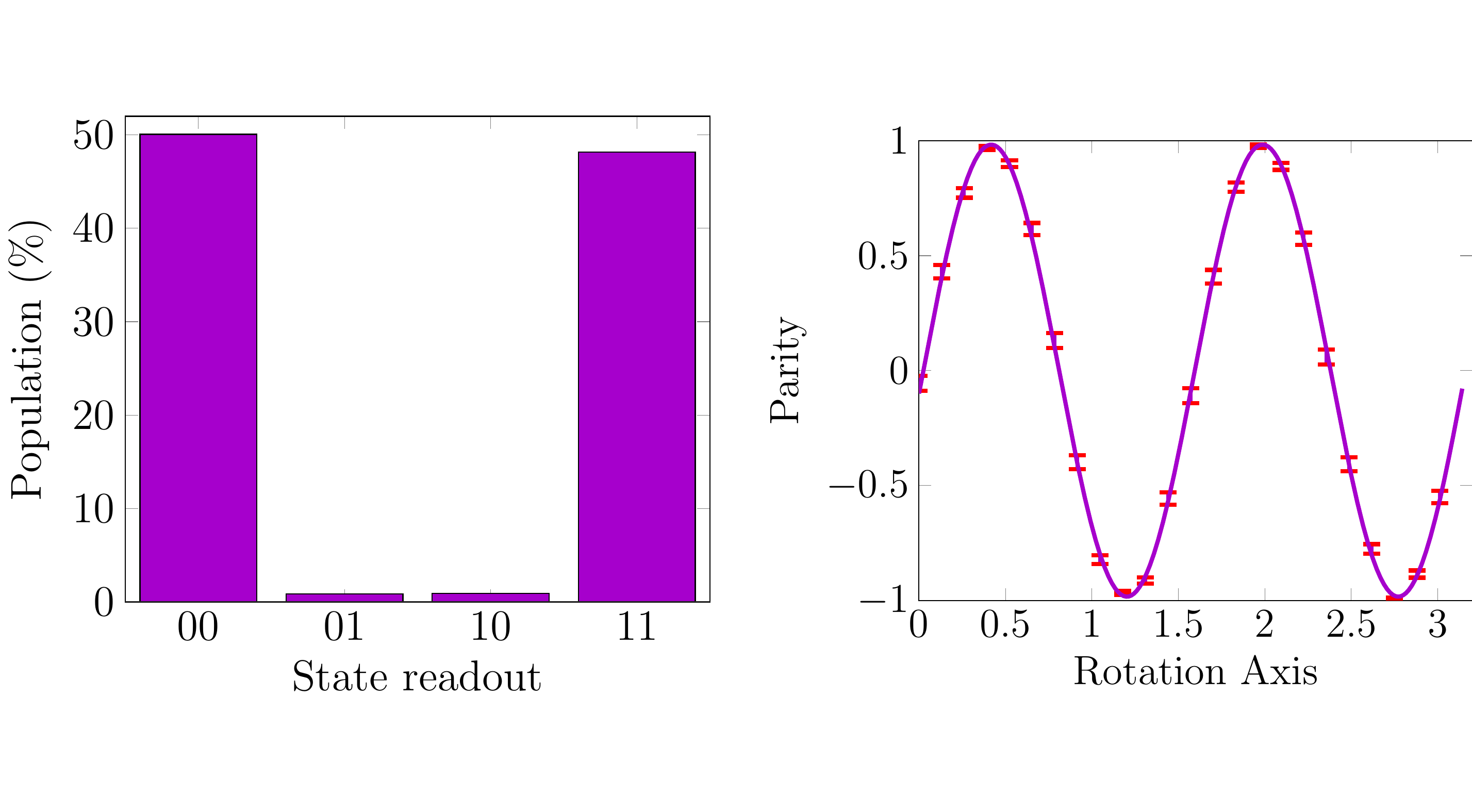}}
	\caption{\small (a) State population and (b) parity scan of the two qubits after the optimized and robust two-qubit gate shown in Fig. 1, indicating a fidelity of 98.3(4)\%.}
\end{figure}

We present the results on entangling two neighboring ions on one edge of the ion chain in the robust case. The output population and parity are measured and shown in Figs. 3(a) and (b), giving a SPAM-corrected fidelity of $98.3(4)\%$. This is among the highest fidelities achieved for multi-qubit gates in the presence of spectator ions \cite{small_computer}. Using the robust gate, we also successfully perform a CNOT gate with 98.6(7)\% fidelity and generate a 3-qubit GHZ state with 92.6(3)\% fidelity, whose results are demonstrated in the Supplemental Material.

In order to lower the overall laser intensity $\Omega$, each 90 \textrm{$\mu$}s pulse is performed twice for each gate, with a combined gate time of 180 \textrm{$\mu$}s. The $\Omega$ required is $2\pi\times600$ kHz in carrier Rabi frequency, which is much larger than $2\pi\times151$ kHz as expected by simulation. The discrepancy is most likely due to an overestimate of the Lamb-Dicke parameters in our simulation. The high power used worsens other error sources such as Raman scattering, off-resonant excitation, and crosstalk with other qubits \cite{Lucas2, HF_Wineland}, which may contribute to the 1\% error level observed.

The theoretically estimated gate error is plotted as a function of frequency drift in Fig. 4(a) to compare the robust pulse with non-robust. A typical error threshold for high-fidelity gates is 0.01\%. The robust pulse can tolerate frequency errors up to $\pm1.5$ kHz, whereas the non-robust less than $\pm 0.1$ kHz. The non-robust pulse has a quadratic dependence on the drift, whereas the robust version has a quartic dependence. This is expected, since error is proportional to displacement squared, and the first-order dependence of the displacement on drift is canceled out in the robust case.

\begin{figure}
	\scalebox{0.8}{\includegraphics{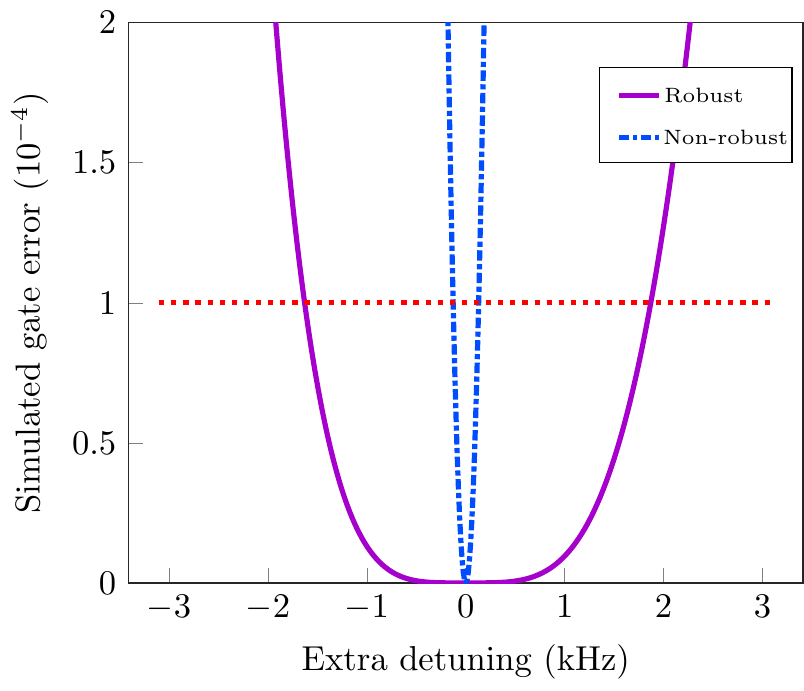}}
\end{figure}
\begin{figure}
	\scalebox{0.8}{\includegraphics{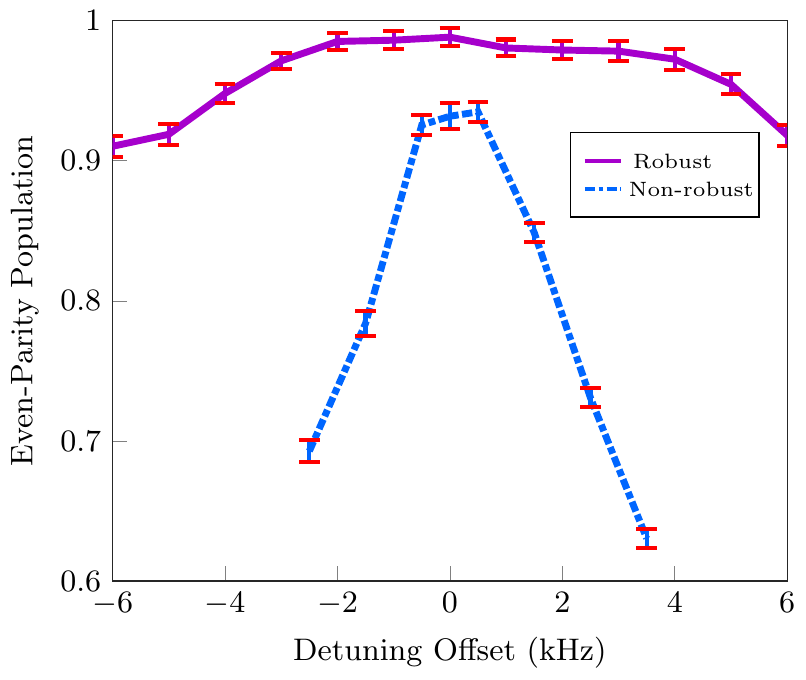}}
	\caption{\small (a) Simulated gate error and (b) Experimental even-parity populations of the two qubits after the gate for a range of detuning offsets. The robust gate has a significantly better performance than non-robust in both theory and experiment}
\end{figure}

To determine the impact of sideband drifts, we experimentally run the two gates over a range of symmetric detuning offsets (Fig. 4(b)). The robust version has even-parity population higher than 90\% for frequency offsets up to $\pm5$ kHz, whereas the non-robust gate has significantly lower fidelity and tolerance towards frequency errors (within $\pm1$ kHz), confirming that the robust method improves fidelity significantly by canceling errors due to frequency drifts.  \smallskip

\PlaceText{17mm}{25mm}{\small (a)}
\PlaceText{61mm}{25mm}{\small (b)}

\PlaceText{113mm}{19mm}{\small (a)}
\PlaceText{113mm}{80mm}{\small (b)}

To test the scalability of our method, we run a similar optimization for 17 ions, motivated by the 17-qubit surface code proposed for quantum error correction \cite{qec_new, qec1,qec2,qec3}. The sideband frequencies are calculated from a simulated anharmonic ion trap with an average ion separation of about 3.5 $\mu$m. Such high ion density may be challenging to realize with current technology, but that does not pose a fundamental physical limit to experiments.

\begin{figure}
	\centering
	\scalebox{0.9}{\includegraphics{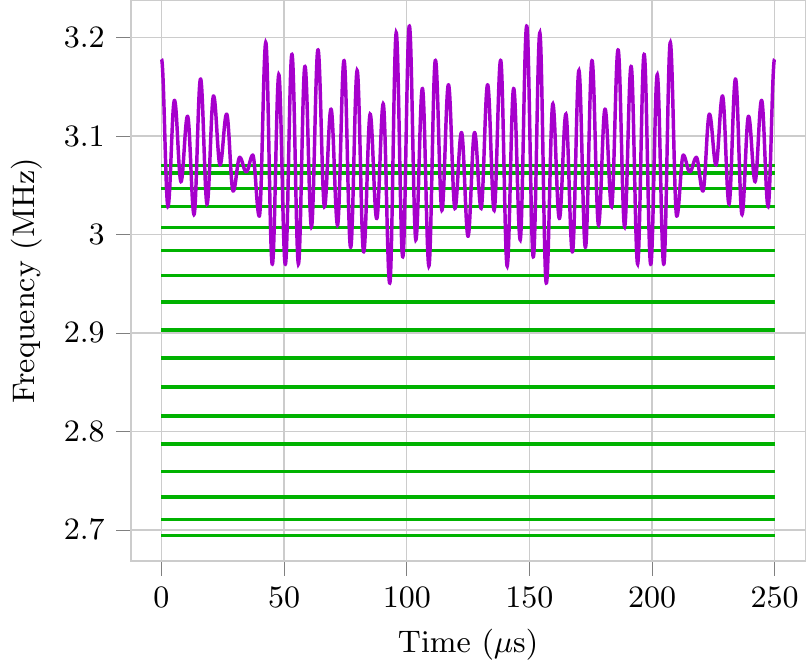}}
	\caption{\small Optimized FM two-qubit gate for 17 ions. The sideband frequencies (green) are obtained by simulations}
\end{figure}

The robust FM pulse obtained consists of 47 oscillations within a gate time of 250  $\mu$s (Fig. 5). The gate can tolerate a frequency drift of 500 Hz for an error threshold of 0.01\%. Apparently, the gate is more sensitive to frequency errors due to an increased number of motional modes and a longer gate time.

The power required ($\Omega$) for the two-qubit gate ranges from $2\pi \times 115$ kHz for neighboring ions to $2\pi\times249$ kHz for the furthest separated ions ($\approx$ 1:2 ratio between lowest and highest). This is an encouraging result. Previous simulation results indicate that two-qubit gate time and power increase very quickly with the distance between the ions. But by using a flexible and well-designed optimization program, we have found an FM pulse that can overcome this difficulty.

We have shown that we can perform high-fidelity two-qubit gates in a 5-ion trap using frequency modulation. In theory, the optimized robust FM pulse can suppress errors in gate fidelities to below $0.01 \%$ for up to a $\pm$1.5 kHz frequency offset for 5 $^{171}$Yb$^+$ ions. The gate is used to maximally entangle two ions in experiment and has a fidelity of $98.3(4)\%$. We speculate that in the near future, we will attain over 99.9\% fidelity previously achieved with 2-ion chains \cite{Lucas1, Lucas2, HF_Wineland}.

We would like to thank Todd Green, Luming Duan, and Gang Shu for useful discussions. This work was supported by the Office of the Director of National Intelligence - Intelligence Advanced Research Projects Activity through ARO contract W911NF- 10-1-0231 and the ARO MURI on Modular Quantum Systems.

\onecolumngrid
\section{\large Supplemental Material}
\vspace{5mm}
\twocolumngrid
 
\section{Additional experimental results}

\begin{figure}[h]
	\centering
	\scalebox{0.3}{ \includegraphics{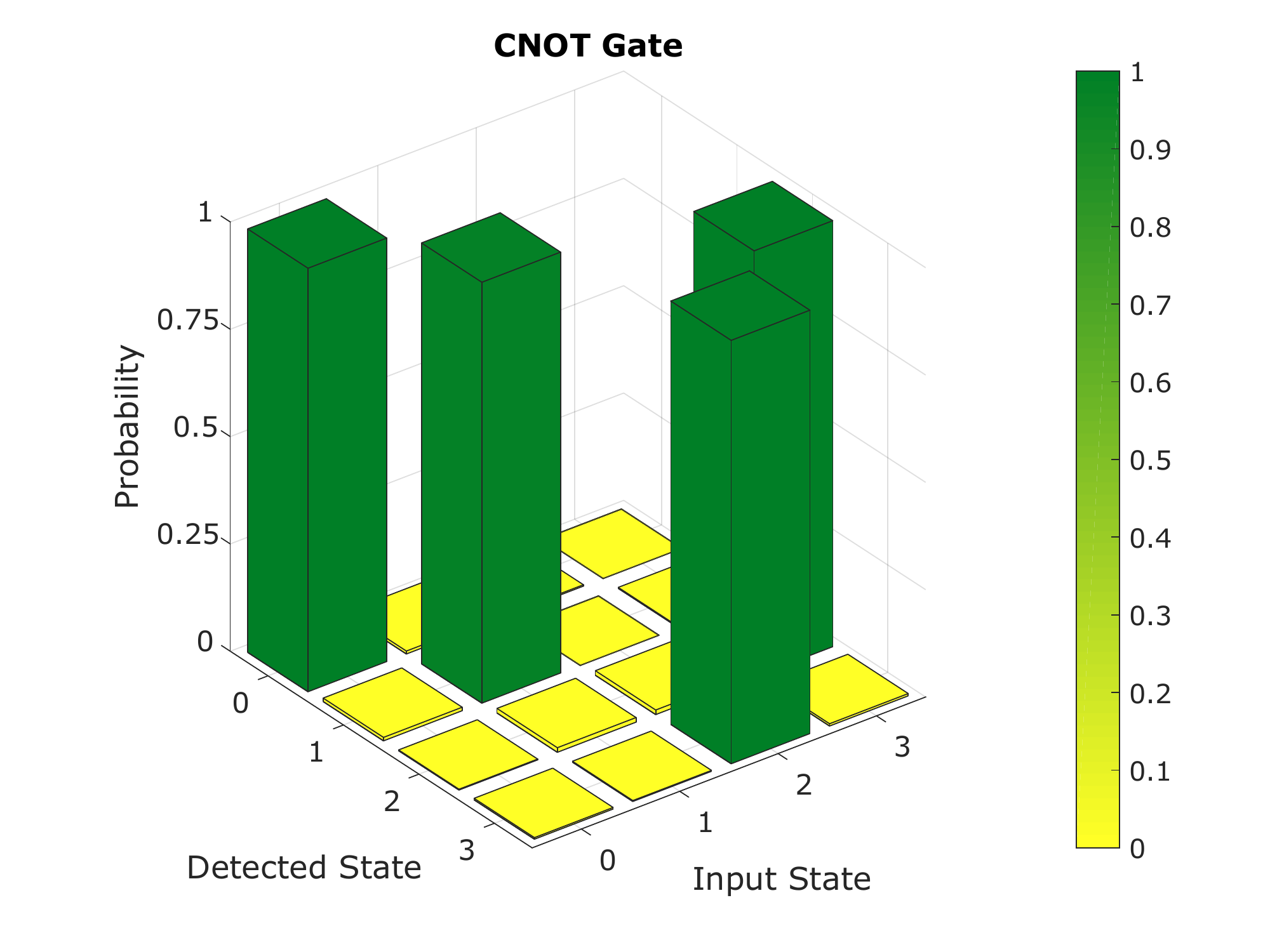}}
	\caption{The probability of each output state for any input state after the CNOT gate, with fidelity 98.6(7)\%}
\end{figure}

\begin{figure}[h]
	\centering
	\scalebox{0.19}{\includegraphics{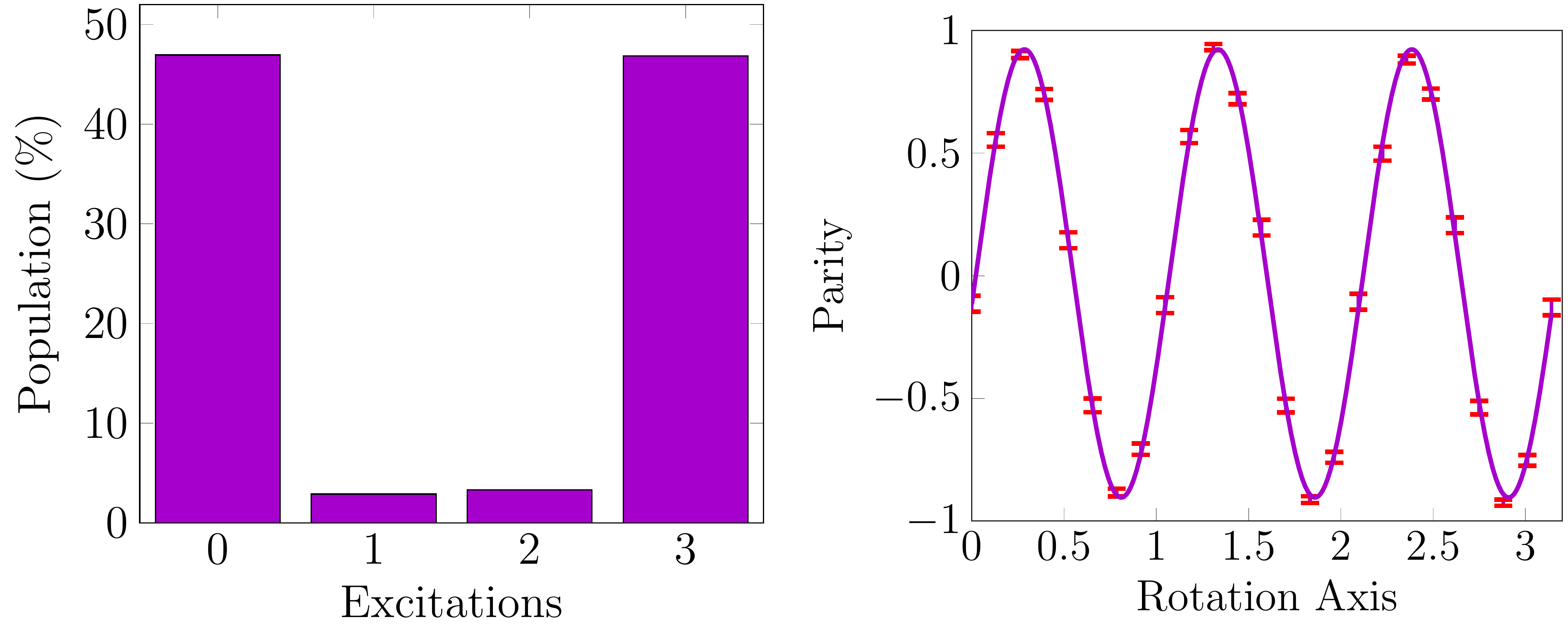}}
	\caption{(a) Combined population for any number of excitation and (b) parity scan of the output 3-qubit GHZ state, giving a fidelity of 92.6(3)\%}
\end{figure}

\begin{figure}[h]
	\begin{subfigure}
	\centering
	\scalebox{0.8}{\includegraphics{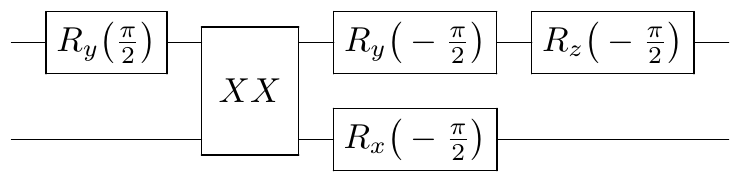}}
	\vspace{5mm}
	\end{subfigure}

	\begin{subfigure}
	\centering
	\scalebox{0.8}{\includegraphics{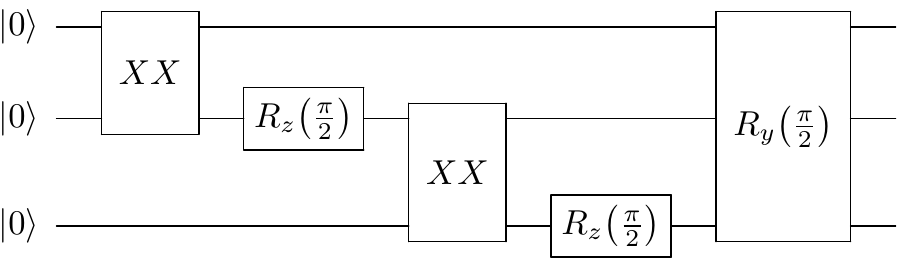}}
	\end{subfigure}
	\caption{Circuit diagrams for (a) the CNOT gate and (b) the generation of a 3-qubit GHZ state. XX stands for our FM two-qubit entangling gate and $R_x$, $R_y$, and $R_z$ stand for single-qubit rotations.}
\end{figure}

\PlaceText{17mm}{110mm}{\small (a)}
\PlaceText{61mm}{110mm}{\small (b)}

\PlaceText{15mm}{164mm}{\small (a)}
\PlaceText{15mm}{185mm}{\small (b)}

We experimentally perform a CNOT gate with a fidelity of 98.6(7)\% (Fig. 6), using one robust FM two-qubit entangling gate and several single-qubit gates (Fig. 8(a)). We also successfully create a 3-qubit GHZ state with a fidelity of 92.6(3)\% (Fig. 7), using two robust two-qubit gates and several single-qubit gates (Fig. 8(b)). These results give further proof that the FM two-qubit gate is a working tool for quantum logic operations.

\section{M\o lmer-S\o rensen gate for varying detuning}

This section reviews the physics of a standard M\o lmer-S\o resen gate. Note that the most important generalization made in this paper is the time-dependence of detuning. The laser phase must be kept continuous, which should be more easily achieved in experiments than otherwise.

Suppose the driving field consists of two counter-propagating laser beams with the same intensity and opposite detunings, applied to any two ions in a linear N-ion crystal. We assume that the beams are perpendicular to the ion chain axis, so that only the transverse motional modes are excited. The ion-field interaction can be written as \cite{Molmer, Sorensen}:

\begin{equation}
\hat{H}_{MS}
= \dfrac{\Omega}{2} \sum\limits_{k=1}^{N} S^k_{\phi,\gamma} a_k^{\dagger} e^{i\theta_k(t)} + {S^k_{\phi,\gamma}}^{\dagger} a_k e^{-i\theta_k(t)}\\
\end{equation}

where $\theta_k$ is the integrated phase of the detuning between the driving force and the k-th sideband, i.e. $\theta_k(t) = \int_{0}^{t}\delta_k(t')dt'$, and $\Omega$ is the effective Rabi frequency for the carrier transition using a particular laser intensity. $S^k_\phi$ equals $\eta_{i,k}\sigma_\phi^i + \eta_{j,k} e^{i\gamma} \sigma_\phi^j$, where $\sigma_\phi = \sigma_x\cos{\phi}+\sigma_y\sin{\phi}$ is a general spin flip operator about an axis on the x-y plane, $\phi$ is half the relative phase between the two sidebands, and $\gamma$ is the relative phase between the lasers applied to the two ions. $\eta_{j,k}$ is the Lamb-Dicke parameter for the $j$th ion and the $k$th motional mode, and is given by $\Delta k\sqrt{\dfrac{\hbar}{2m\omega_k}} u_{jk}$, where $\Delta k = 4\pi/\lambda$ is the wavenumber of the two counterpropagating Raman lasers ($\lambda$ = 355 nm), and $u_{jk}$ is the unitary matrix that maps ion coordinates to the resonant mode coordinates. Note that if the lasers are at an angle to the axis of motion, the parameter will be reduced by the cosine of that angle. The expression is valid if the Lamb-Dicke approximation holds ($\eta_{j,k}\sqrt{n+\frac{1}{2}} \ll 1, \hspace{5pt} n = \sqrt{\langle a^\dagger a\rangle}$), and the direct carrier transition is small ($\Omega$ is much smaller than the detuning from the carrier transition).

The Hamiltonian consists of a sum of products of internal and motional operators, and thus represents a state-dependent force acting on the ion chain as a whole. To solve the time-dependent Schr\"{o}dinger equation, we apply the Magnus expansion to compute the argument of the effective propagator \cite{AM1, Roos}:

\begin{equation}
\ket{\psi(t)} =  \hat{D} (\{\hat{\alpha}_k\}) \hat{E}(\beta_{ij}) \ket{\psi(0)} \\
\end{equation}
\begin{equation}
\begin{aligned}
\hat{D} (\{\hat{\alpha}_k\}) &= \exp(\sum\limits_{k=1}^N  (\hat{\alpha}_k a_k^{\dagger} - \hat{\alpha}_k^\dagger a_k)) \\
&\text{where} \hspace{5pt} \hat{\alpha}_k(t) = S^k_{\phi,\gamma}\dfrac{\Omega}{2} \int^{t}_0 e^{i\theta_k(t')}dt' \\
\end{aligned}
\end{equation}
\begin{equation}
\begin{aligned}
\hat{E}(\beta_{ij}) &= \exp(-i\beta_{ij}\sigma_\phi^i\sigma_\phi^j) \\
&= \exp \biggl(-i\sigma_\phi^i\sigma_\phi^j\dfrac{\Omega^2}{2}\cos{\gamma}\sum\limits^N_{k=1}\int^{t}_0 \int^{t'}_0\eta_{i,k}\eta_{j,k} \\
&\times\sin(\theta_k (t')-\theta_k (t'')) dt'dt''\biggr)
\end{aligned}
\end{equation}

The first term from the expansion is the direct time integral of the Hamiltonian and is proportional to $ \hat{\alpha}_k a_k^{\dagger} - \hat{\alpha}_k^\dagger a_k$, which is the argument of the displacement operator and is related to quantum coherent states. The second term is the double time integral of the commutator of the Hamiltonian as functions of different time parameters, and is proportional to $\sigma_\phi^i\otimes\sigma_\phi^j$. Conveniently, higher-order terms vanish, and the two surviving terms commute, so we can express the final propagator as the product of two unitaries.

Consider the first operator $\hat{D} (\{\hat{\alpha}_k\})$, where the displacement $\hat{\alpha}_k$ is state-dependent and is proportional to the spin operator $S^k_{\phi,\gamma}$. If the internal state happens to be an eigenstate of $S^k_{\phi,\gamma}$, we may replace it with its eigenvalue, and $\hat{D}$ simply displaces the motional state from one coherent state to another by $\alpha_k$. We can plot the 2-D phase space trajectory (PST) to keep track of the complex displacement over time. It is worth emphasizing that the quadrature axes in the PSTs do not represent the expected position or momentum of any particle like they do for a single quantum harmonic oscillator, since we are looking at the Hamiltonian in the interaction frame, and we are tracking down the collective instead of individual motion of the ions.

In general, the initial internal state is a superposition of the four eigenstates of $S^k_{\phi,\gamma}=\eta_{i,k}\sigma_\phi^i + \eta_{j,k} e^{i\gamma} \sigma_\phi^j$, and each eigenstate follows a different trajectory in the phase space according to its eigenvalue. For tidiness, we only track the trajectory of $\ket{++}_\phi$, where $\ket{+}_\phi$ is the positive eigenstate of $\sigma^i_\phi$, in the case where the laser phase $\gamma$ is zero. Since the trajectories for different eigenstates have different end points, there is a residual entanglement between the internal and motional state spaces, which will result in a mixed internal state since we do not measure the ion motion. Thus, we need $\alpha_k(t) = 0$ in magnitude for all motional modes $k$ in our optimization to guarantee that end points of the trajectories are sent back to their starting points.

The second operator $\hat{E}_{ij}$ represents a rotation on the Bloch sphere spanned by $\ket{\downarrow\downarrow}$ and $\ket{\uparrow\uparrow}$. For maximal entanglement we set $\gamma$ = 0 and require the magnitude of the argument of the exponential to be $\pi/4$ to effect a $\pi/2$ rotation, which maps $\ket{\downarrow\downarrow}$ to $\frac{1}{\sqrt{2}}(\ket{\downarrow\downarrow}+i e^{2i\phi}\ket{\uparrow\uparrow})$. We simply adjust $\Omega$ to satisfy this requirement since it is a free constant parameter. If $\Omega$ is too large, we repeat the gate sequence $R$ times to lower it by a factor of $\sqrt{R}$. We may also alter the axis of rotation by changing phase lag between the sidebands $\phi$.

\section{Error estimate due to spin-motion entanglement}

This section gives a simplified justification for the error estimate presented in equation (3) in the main text.

Suppose the internal state is an equal superposition between $\ket{\Psi_1}$ and $\ket{\Psi_2}$, which are eigenstates of some spin operator $\hat{S}$, with eigenvalues $\pm 1$. The system is subject to the effect the displacement operator $\hat{D} (\alpha) = \exp(\hat{S}(\alpha a^\dagger - \alpha^* a))$, so the two eigenstates have opposite displacements $\pm \alpha$ from the origin, which is ideally zero. Assuming that the ions are perfectly cooled to the ground state (a reasonable approximation for this experiment), the final and ideal states will be:

\begin{equation}
\begin{aligned}
\ket{\psi_{final}} &= \dfrac{1}{\sqrt{2}}(\ket{\Psi_1,\alpha}+\ket{\Psi_2,-\alpha}) \\
\ket{\psi_{ideal}} &= \dfrac{1}{\sqrt{2}}(\ket{\Psi_1,0}+\ket{\Psi_2,0})
\end{aligned}
\end{equation}

The gate fidelity is given by:
\begin{equation}
\begin{aligned}
|\bra{\psi_{final}}\ket{\psi_{ideal}}|^2 &= \Big|\frac{1}{2}(\bra{\alpha}\ket{0}+\bra{-\alpha}\ket{0})\Big|^2\\
&=e^{-|\alpha|^2} \approx 1-|\alpha|^2
\end{aligned}
\end{equation}

Alternatively, we can trace the associated density matrix $\ket{\psi_{final}}\bra{\psi_{final}}$ over the motional space. By realizing that $\text{tr}(\ket{\alpha}\bra{-\alpha})=\text{tr}(\ket{-\alpha}\bra{\alpha}) = e^{-2|\alpha|^2}$, in the eigenbasis $\{\ket{\Psi_1},\ket{\Psi_2}\}$, the final density matrix is:
\begin{equation}
\rho_f = \dfrac{1}{2}
\begin{bmatrix}
1       &  e^{-2|\alpha|^2} \\
e^{-2|\alpha|^2}       & 1

\end{bmatrix}
\end{equation}
And we arrive at the same fidelity:
\begin{equation}
F = \bra{\psi_{ideal}}\rho_f\ket{\psi_{ideal}} \approx 1-|\alpha|^2
\end{equation}
Since there are multiple motional modes for a multi-ion chain, the total error is simply the sum of $|\alpha|^2$ for all modes.

The motional displacement is difficult to determine since it is inherently state-dependent, and the initial state is assumed to be arbitrary. By observing the original expression for $\hat{\alpha}_k(t)$, we approximate the error as:
\begin{equation}
|\alpha_k| \approx \tilde{\eta}\tilde{\Omega}\bigg|\int^{\tau}_0 e^{i\theta_k(t)}dt\bigg|
\end{equation}
where $\tilde{\eta} = \eta_{j,0} = \Delta k\sqrt{\dfrac{\hbar}{2m\omega_x}}\dfrac{1}{\sqrt{N}}$ is the Lamb-Dicke parameter for all ions for the common mode (0.047 for 5 $^{171}$Yb$^+$ ions, 0.025 for 17 ions), and $\tilde{\Omega}$ is the approximate power required to entangle a pair of qubits (about $2\pi\times200$ kHz). Thus we define the gate error $\varepsilon$ to be:
\begin{equation}
\begin{aligned}
\bm{\varepsilon} &\approx \sum_{k=1}^{N}|\alpha_k|^2  \\
&\approx (\tilde{\eta}\tilde{\Omega})^2\sum_{k=1}^{N}\bigg|\int^{\tau}_0 e^{i\theta_k(t)}dt\bigg|^2
\end{aligned}
\end{equation}
where $\tilde{\eta}$ is the characteristic size of the Lamb-Dicke parameter, and $\tilde{\Omega}$ is the approximate power required to induce maximum entanglement between the qubit pair. Together, $\tilde{\eta}\tilde{\Omega}$ is the overall ``sideband coupling strength", which is approximated as $2\pi\times10$ kHz for 5 ions and $2\pi\times5$ kHz for 17 ions. 

\section{The robustness condition}

This section explains the robustness condition presented in equation (4) in the main text.

Since $\alpha_k \sim \int^{t}_0 e^{i\theta_k(t')}dt' = 0$ is a necessary condition for guaranteeing zero displacement in the motional state space, we investigate how we can suppress $\alpha_k$ up to the first order in $\delta_1$. Replacing the phase $\theta_k$ with $\theta_k + \delta_1 t$, we evaluate the displacement through integration by parts:

\begin{equation}
\begin{aligned}
\alpha_k(t) &\sim \int^{\tau}_{0}e^{i\theta_k (t)+i\delta_1 t}dt \approx \int^{\tau}_{0}(1+i\delta_1 t)e^{i\theta_k (t)}dt \\
&= i\delta_1 \int^{\tau}_{0}t e^{i\theta_k (t)}dt \\
&= i\delta_1\Big( \Big[ t \int^{t}_{0}e^{i\theta_k (t')}dt' \Big]^{\tau}_{0} - 
\int^{\tau}_{0} \int^{t}_{0}e^{i\theta_k (t')}dt'dt \Big)\\
&= i\delta_1\Big( 0 - \tau \alpha_{k,avg} \Big) \\
\end{aligned}
\end{equation}
where $\alpha_{k,avg}$ is the time-averaged position of the trajectory from t = 0 to t = $\tau$.  Therefore, we need it to lie on the starting point in order for $\alpha_k$ to remain zero up to the first order of the drift or uncertainty. Note that the approximation $e^{i\theta_k (t)} \approx 1+i\delta_1 t$ is valid only when $\delta_1 \ll 1/\tau$. Hence, the longer the gate time, the less robust the gate becomes.

In addition, if the pulse is time-symmetric (i.e. $\delta_k(t) = \delta_k(\tau-t)$), the center of mass lying at the origin automatically guarantees that the end point will lie there as well. Thus, the robustness condition ($\alpha_{k,avg} = 0$) is a sufficient condition for displacement minimization ($\alpha_k = 0$) as long as we are restricted to symmetric pulses. The optimization criterion is now simply: 

\begin{equation}
\begin{aligned}
\alpha_{k,avg} \sim \int^{\tau}_{0} \int^{t}_{0}e^{i\theta_k (t')}dt'dt = 0, \hspace{10pt} k = 1,..,N
\end{aligned}
\end{equation}

We seek to vary the detuning during that gate such that the above condition is satisfied. Given sufficient degrees of freedom and a good initial guess, we can arrive at an optimal pulse deterministically.

\section{Optimization process}

Modifying the frequency allows us to alter the trajectories' curvature, and hence their end points and time-averaged positions. In our optimization, we choose the vertices of the frequency oscillations to be our control parameters. The number of vertices correspond to the degrees of freedom needed to achieve an optimal solution. It increases linearly with the number of motional modes. We connect these vertices using the cosine function to create a smoothly varying frequency pattern. This may be a useful feature, since it is difficult to vary physical parameters discretely in real experiments. The overall change in frequency ($\sim$100 kHz) is small compared to the frequencies used by conventional optical modulators ($\sim$100 MHz), minimizing sudden physical changes.

The average frequency lies above all motional modes (blue detuned), but the frequency crosses several sidebands and becomes red detuned with respect to them. The phonon number does not increase dramatically since the driving frequency only overlaps with the sidebands momentarily.

To search for a robust frequency pattern, we define the cost function as the sum of distance squared between the center of mass of each trajectory and its starting point: 
\begin{equation}
\begin{aligned}
\text{Cost} &= \sum\limits_{k=1}^N\hspace{5pt}\bigg| \int^{\tau}_{0} \int^{t}_{0}e^{i\theta_k (t')}dt'dt \bigg|^2 \\
&= \sum\limits_{k=1}^N\bigg(\int^{\tau}_{0} \int^{t}_{0}\cos\theta_k (t')dt'dt\bigg)^2 \\
&+ \bigg(\int^{\tau}_{0} \int^{t}_{0}\sin\theta_k (t')dt'dt\bigg)^2
\end{aligned}
\end{equation}
Similarly, we can define the error for a non-robust pattern as the distance squared between the trajectory end points and starting points.

It is worth noting that the optimization algorithm is inherently deterministic and requires little computational resources. For 5 sidebands, given a good initial guess, we can arrive at an optimal FM pattern in about 30 seconds using a regular laptop computer. \\
\\

\section{Area enclosed by the trajectory}

\begin{figure}[h]
	\centering
	\scalebox{1.2}{\includegraphics{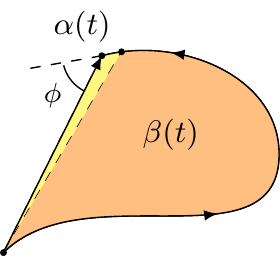}}
	\caption{\small An arbitrary trajectory in complex space}
\end{figure}

This section shows that the area enclosed by a trajectory has a simple expression as a double integral. Consider the following integral:

\begin{equation}
\begin{aligned}
\alpha(t) = \int^{t}_0 e^{i\theta(t')}dt', \hspace{10pt}\theta(t) = \int_{0}^{t}\delta(t')dt',
\end{aligned}
\end{equation}
which is a general representation of a trajectory in the complex plane (see Fig. 9). At any given time $t$ it moves at angular rate $\delta(t)$, angle $\theta(t)$, and speed 1. The area enclosed from $t$ to $t+\delta t$ (yellow triangle in figure) is given by:

\begin{equation}
\begin{aligned}
\frac{1}{2}&\big|\alpha(t)\big|dt\sin(\phi) \\
= \frac{1}{2}&dt \Im\big(e^{i\theta(t)}\alpha^*(t)\big) \\
= \frac{1}{2}&dt \Im\Bigg(\int^{t}_{0}e^{i\theta(t)-i\theta(t')}dt'\Bigg) \\
= \frac{1}{2}&dt \int^{t}_{0}\sin\big(\theta(t)-\theta(t')\big)dt'
\end{aligned}
\end{equation}

Hence the total area enclosed by the trajectory over a period of time $t$ is given by:
\begin{equation}
\begin{aligned}
\beta(t) = \frac{1}{2}\int^{t}_0 \int^{t'}_0\sin\big(\theta(t')-\theta (t'')\big) dt''dt',
\end{aligned}
\end{equation}
This double integral coincides with the entanglement between two qubits after the MS gate, or rather the angle of rotation between $\ket{\downarrow\downarrow}$ and $\ket{\uparrow\uparrow}$. Hence we may evaluate how much entanglement is generated by the MS gate by observing the sizes and shapes of the PSTs.



\end{document}